\documentclass[fleqn,twoside]{article}
\usepackage{espcrc2}
\usepackage{graphicx}

\newlength{\figwidth}
\renewcommand{\vec}[1]{\mbox{\boldmath $#1$}}
\renewcommand{\Re}{\mathop{\rm Re}}
\newcommand{\Tr}{\mathop{\rm Tr}}

\title{Numerical study of staggered fermion on anisotropic lattices%
  \thanks{Talk presented by K. Nomura}}

\author{
 Kouji Nomura%
  \address{Department of Physics, Hiroshima University, 
           Higashi-Hiroshima 739-8626, Japan \vspace{-0.25cm}},
 Takashi Umeda%
  \address{Yukawa Institute for Theoretical
           Physics, Kyoto University, Kyoto 606-8502, Japan},
  and
 Hideo Matsufuru$^{\rm b}$,}

\begin{document}

\begin{abstract}
We study calibration procedures of the staggered quarks on
anisotropic lattices in the quenched approximation and
in $N_f = 2$ dynamical simulations.
For the calibration conditions we adopt the hadronic radii $r_0$
and the meson masses in the temporal and spatial directions.
On the quenched lattice, we calibrate the quark field and compare
the result with the result determined using the meson dispersion
relation.
In dynamical simulations, we determine the anisotropy parameters
$\gamma_G$ and $\gamma_F$ simultaneously within  1\% accuracy
at renormalized anisotropy $\xi = a_s/a_t = 4$.
\vspace{1pc}
\end{abstract}

\maketitle

\section{Introduction}
Anisotropic lattices realize small temporal lattice spacings
while keeping modest computational costs.
The technique is useful in various fields of lattice QCD simulation:
At finite temperature the large number of degrees of freedom
in Euclidean time direction leads large number of Matsubara
frequencies, which is efficient for calculation of the equation of
state \cite{EOS} and for analyses of temporal correlation functions
\cite{TARO,Charm,Gball2}.
The large temporal cutoff is important for
relativistic formulations of heavy quarks on lattices \cite{heavy}.
It is also convenient for the correlators in which noises
quickly grow as time such as glueballs \cite{Gball2,Gball1}
and negative parity baryons \cite{NPbaryon}.

However, uncertainties of anisotropy parameters bring additional
errors into the observed quantities.
For precise calculations, we need to tune anisotropy parameters 
with good statistical accuracies and to control the systematic
errors in the continuum extrapolations.
In this work we discuss the tuning procedures of anisotropy
parameters for the staggered fermions in the quenched and dynamical
QCD simulations.
Our goal is to tune anisotropy parameters within 1\% statistical
errors at each set of $\beta$ and $m_0$.
In the following, we focus on lattices with the renormalized 
anisotropy $\xi = a_s/a_t = 4$.


\section{Lattice action}
In the numerical simulations, we adopt
Wilson gauge and staggered quark actions on anisotropic lattices.
The Staggered quarks have several advantages over the Wilson-type
quarks in studies related to the chiral symmetry
and in simulations at lighter quark masses.
The gauge action is
\begin{eqnarray}
S_G & = & \beta \sum_x \left\{ \sum_{i<j}^{3} \frac{1}{\gamma_G}
 \left[ 1-\frac{1}{3}\Re \Tr  U_{ij}(x) \right] + \right.
 \nonumber \\
    &   & \left.  \sum_{i}^{3} \gamma_G
 \left[ 1-\frac{1}{3}\Re \Tr  U_{i4}(x) \right] \right\} ,
\end{eqnarray}
where $\gamma_G$ is the bare gauge anisotropy parameter.
The quark action is defined as   
\begin{eqnarray}
S_F  =  \sum_{x,y} \bar{\chi}(x) K(x,y) \chi(y).  
\end{eqnarray}
\begin{eqnarray}
K(x,y)  = \delta_{x,y} & & \nonumber \\ 
& &\hspace{-29mm}
   - \gamma_F\kappa_\sigma
   \eta_4(x)\left[  U_4(x)\delta_{x,y-\hat{4}}-
              U^\dagger_4(x-\hat{4})
                           \delta_{x,y+\hat{4}}\right] \nonumber\\
& &\hspace{-29mm}
   - \kappa_\sigma \sum_i
   \eta_\mu(x) \left[ U_i(x)\delta_{x,y-\hat{i}}-
              U^\dagger_i(x-\hat{i})
                           \delta_{x,y+\hat{i}}\right],
\end{eqnarray}
where $\gamma_F$ is a bare fermionic anisotropy parameter,
$\eta_\mu(x)$ the staggered phase,
$\kappa_{\sigma}= 1/2m_0$ with $m_0$ the bare quark mass in spatial
lattice units.

Although these anisotropic lattice actions have been adopted
in Ref.\cite{DynamicalAniso}, they did not discuss the
systematic uncertainties due to the anisotropy to the sufficient
level for precision computations.

\section{Calibration procedures}

The anisotropy parameters $\gamma_G$ and $\gamma_F$ should be tuned
so that the physical isotropy condition,
$\xi_G(\gamma_G^*,\gamma_F^*)=\xi_F(\gamma_G^*,\gamma_F^*)=\xi$ 
holds at each set of $\beta$ and $m_0$,
where $\xi_G(\gamma_G,\gamma_F)$ and
$\xi_F(\gamma_G,\gamma_F)$ are the renormalized anisotropies
defined through the gauge and quark observables, respectively.
In quenched simulations, one can firstly determine
the $\gamma_G$ independently of $\gamma_F$, and then tune
$\gamma_F$ for fixed $\gamma_G$.
On the other hand, in dynamical simulations, $\gamma_G$ 
and $\gamma_F$ need to be tuned simultaneously.

\subsection{Gauge sector}

For the calibration in the gauge sector,
we define $\xi_G(\gamma_G,\gamma_F)$ through the hadronic radii
$r_0$ \cite{Sommer} in the spatial and temporal directions.
Since we set the lattice scale via $r_0$, the physical anisotropy
is kept in the continuum extrapolation.
This is an advantage to avoid that the systematic errors brought
by the anisotropic lattice remain in the continuum limit.

The value of $r_0$ is defined with the static $q$-$\bar{q}$ potential
through the relation  $r_0^2 F(r_0)=1.65$, where $F(r)$ is 
the force between quarks.
Then $\xi_G$ is defined with the ratio of $r_0$'s in the temporal and
spatial directions: $\xi_G = r_0^{(t)}/r_0^{(x)}$.

In quenched simulations, we can apply the L\"uscher-Weisz noise reduction
technique \cite{LW} and determine $\xi_G$ at 0.2\% level of accuracy
\cite{MOOU03}.
In dynamical simulations, we instead apply the smearing technique to 
the anisotropic three-dimensional planes which enables
determination of $\xi_G$ at 1\% level of accurately.

\subsection{Quark sector}

In the fermion sector,
we define $\xi_F$ with the ratio of meson masses in temporal and
spatial directions:
$\xi_F = m_{h}^{(z)}/m_{h}^{(t)}$ \cite{TARO}.
We use the pseudoscalar channel, since in that channel the oscillating
components of the correlators disappear quickly and hence
statistical errors of masses are small.

Alternative definition of the $\xi_F$ makes use of the 
meson dispersion relation \cite{CPPACS},
\begin{eqnarray}
E^2(\vec{p})=m^2 +\vec{p}^2/\xi_F^2.
\end{eqnarray}
Here $E(\vec{p})$ and $m$ are in temporal lattice units and
$\vec{p}$ is in spatial lattice units, and hence $\xi_F$ appears.
$p_i=2\pi n_i/L_i$ ($i=$1,2,3), where $L_i$ is lattice size
in $i$-th direction.
We use $\vec{n}=(0,0,0),(0,0,1)$ to define $\xi_F$.

At present stage of this work, we adopt the former procedure,
with the ratio of the temporal and spatial meson masses,
as the main calibration procedure for the quark field,
because of smaller statistical uncertainties than the 
latter definition.
In quenched simulation, we compare the results with these two
procedures.

\section{Numerical results of calibration}

\subsection{Quenched simulation}

\begin{figure}[tb]
\vspace{0.3cm} \hspace{0.3cm}
\includegraphics[width=7.0cm]{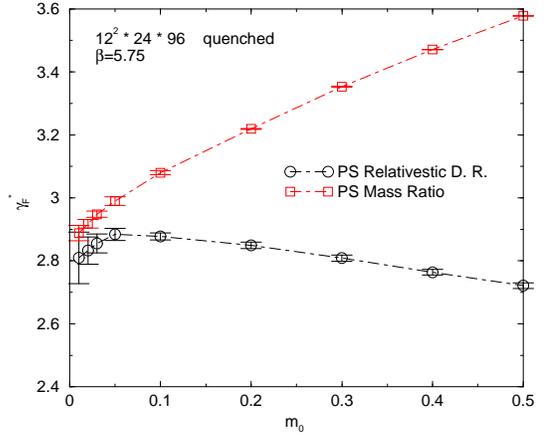}
\vspace{-1.1cm}
\caption{
The result of calibration of the staggered quark field
in the quenched simulation.}
\label{fig1}
\vspace{-0.4cm}
\end{figure}

The simulation is performed on a $12^2 \times 24 \times96$ lattice 
at $\beta=5.75$ and $\gamma_G=3.136$ which correspond to the
spatial cutoff $a_{\sigma}^{-1}= $1.1 GeV and
the renormalized anisotropy $\xi_G=4$ \cite{MOOU03}.
The statistics are 224 configurations.
The bare quark mass range is from 0.01 to 0.5 which 
corresponds to the range $m_{PS}/m_V$ from 0.33 to 0.90.

We calculate the meson mass ratio at several values of $\gamma_F$
and interpolate them to $\gamma_F^*$ at which $\xi_F=4$ holds.
The result is displayed in Fig.~{\ref{fig1}}. 
For our lightest quark mass case, $\gamma_F^*$ is determined
within 0.5\% statistical error.

In Fig.~{\ref{fig1}}, we also show the result of calibration 
using the meson dispersion relation.
The results of two procedures are consistent only in
vary light quark mass region.
To understand the discrepancies of two procedures
and large change of $\gamma_F^*$ from the calibration with mass
ratio at larger quark mass region are subjects of future work.

\subsection{dynamical simulation}

\begin{figure}[tb]
\hspace*{0.2cm}
\includegraphics[width=6.7cm]{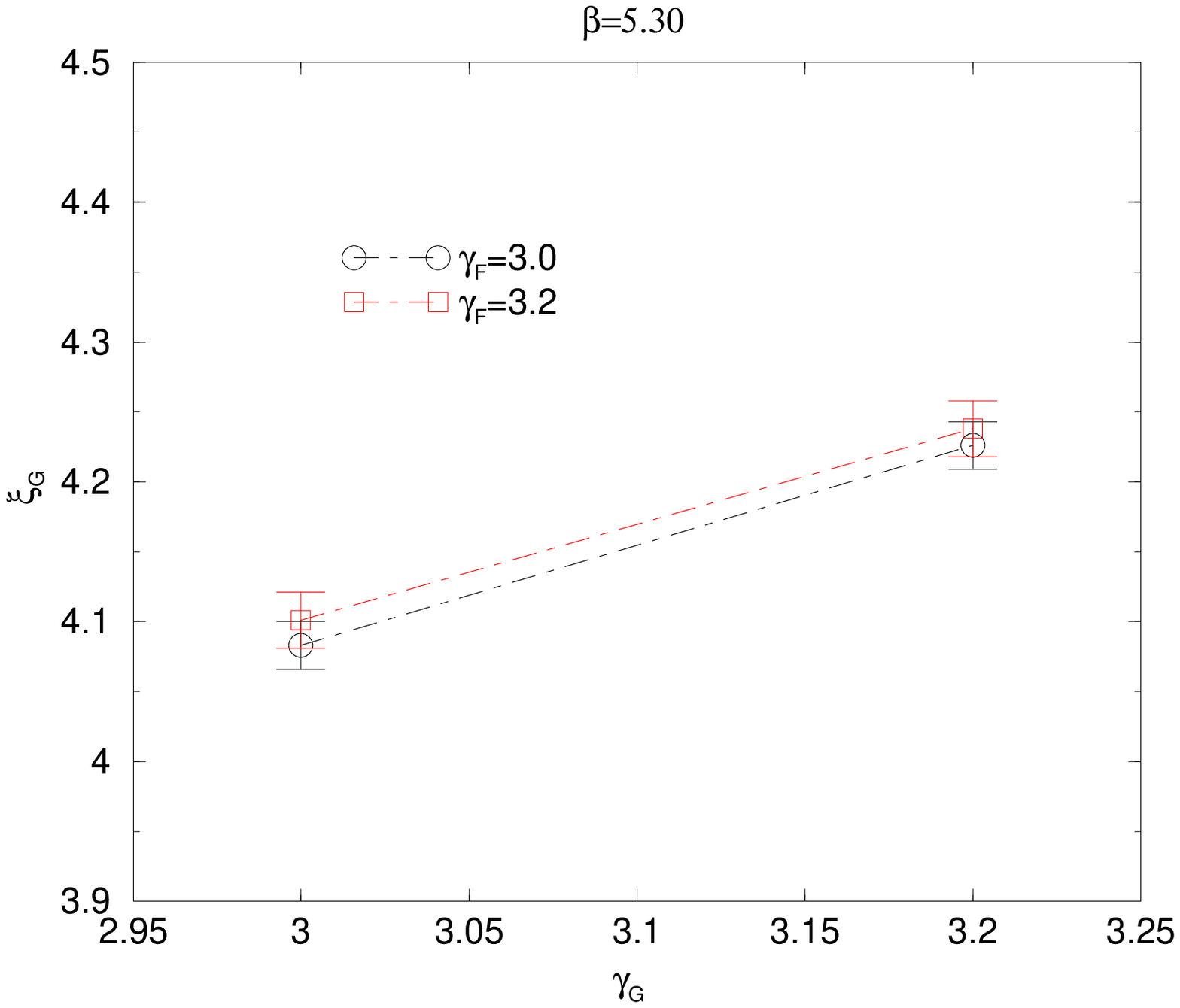}
\vspace{-0.06cm}\\
\hspace*{0.2cm}
\includegraphics[width=6.7cm]{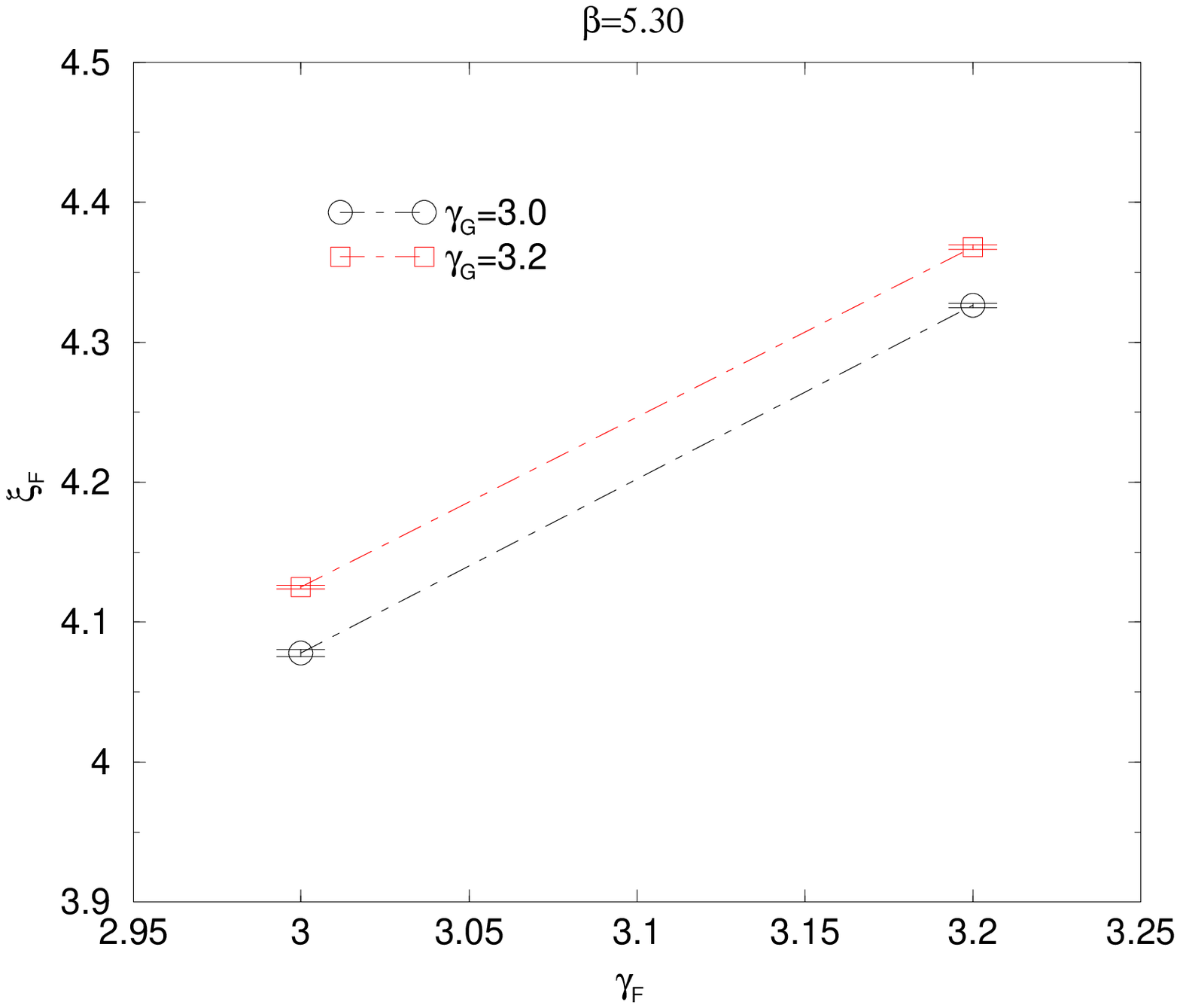}
\vspace{-1.0cm}
\caption{
The result of calibration in the dynamical simulation
at $\beta=5.3$, $m_0=0.1$.
The top and bottom panel show the dependences of
$\gamma_G$ on $\xi_G$ and $\gamma_F$ on $\xi_F$, respectively.}
\label{fig2}
\vspace{-0.5cm}
\end{figure}

The simulation is performed with R-algorithm with $N_f=2$
on lattices of the size $10^2\times 20 \times 80$
at 4 values of $\beta$ from 5.3 to 5.45 and the quark mass $m_0=0.1$.
The statistics are from 300 to 800 trajectories after 200
for thermalization.
In the following we present only the result at $\beta=5.3$ on which 
the spatial cutoff is $a_{\sigma}^{-1}=$ 0.69 GeV and
$m_{PS}/m_V=0.6$.

We determine $\gamma_G^*$ and $\gamma_F^*$ by a linear interpolation
to $\xi = 4$ using the 6 parameter sets of ($\gamma_G$,$\gamma_F$)
assuming the following relations:
\begin{eqnarray}
\xi_G &=&
  a_0 + a_1 (\gamma_G - \gamma_G^*) + a_2 (\gamma_F - \gamma_F^*),
 \nonumber \\
\xi_F &=&
 b_0 + b_1 (\gamma_G - \gamma_G^*) + b_2 (\gamma_F - \gamma_F^*).
\label{eq:xi_linear}
\end{eqnarray}
The result of calibration with the meson mass ratio
is presented in Fig.{\ref{fig2}}.

The $\chi^2$ fit results in
$[a_0,a_1,a_2]=[4.000(24),0.702(92),0.075(92)]$,
$[b_0,b_1,b_2]=[4.000(28),0.220(08),1.225(08)]$.
We find that $\xi_F$ has small dependence on $\gamma_G$
and $\xi_G$ on $\gamma_F$.
By solving Eq.~(\ref{eq:xi_linear}),
we obtain $\gamma_G^* = 2.885(35)$, $\gamma_F^* = 2.955(31)$.
We can tune the parameters $\gamma_G$ and $\gamma_F$ with
the 1\% level of statistical errors.

This calculation was done on a Hitachi SR8000 at KEK.
H.M. and T.U. thank JSPS for Young Scientists for financial support.

\end{document}